\documentclass[twocolumn,showpacs,floatfix,nofootinbib,prc,aps,10pt]{revtex4-1}

\usepackage{amsmath}
\usepackage{amssymb}
\usepackage{dcolumn}
\usepackage{float}
\usepackage{graphicx}
\usepackage{grffile}
\usepackage{hyperref}
\usepackage{multirow}
\usepackage{txfonts}
\usepackage{textcomp}

\newcommand{\eq}[1]{(\ref{#1})}

\newcommand{\feynp}[1]{#1\kern-0.45em/}
\newcommand{\feynq}[1]{#1\kern-0.46em/}
\newcommand{\s}[1]{\sigma_\textsc{#1}}

\newcommand{\GeV}{\;\text{GeV}}
\newcommand{\MeV}{\;\text{MeV}}
\newcommand{\keV}{\;\text{keV}}
\newcommand{\thcm}{\theta^*_K}
\newcommand{\cndf}{\chi^2_\textsc{ndf}}
\newcommand{\fm}{\;\text{fm}}

\newcolumntype{C}[1]{>{\centering\let\newline\\\arraybackslash\hspace{0pt}}m{#1}}
\newcolumntype{L}[1]{>{\left\let\newline\\\arraybackslash\hspace{0pt}}m{#1}}
\newcolumntype{.}{D{.}{.}{-1}}

\renewcommand{\bm}[1]{\boldsymbol{#1}}
\renewcommand{\d}{\mathrm{d}}
\renewcommand{\P}{\mathcal{P}}
\renewcommand{\Lambda}{\varLambda}

\allowdisplaybreaks
\graphicspath{{./figures/}}
\DeclareMathAlphabet{\mathcal}{OMS}{cmsy}{m}{n}
\DeclareMathSymbol{\alpha}{0}{letters}{"0B}
\hypersetup{
colorlinks=true,
linkcolor=blue,
citecolor=blue,
urlcolor=blue,
bookmarksnumbered,
bookmarksopen,
bookmarksopenlevel=1}

\begin{document}

\title{\texorpdfstring{$K^+\Lambda$}{K+ Lambda} electroproduction above the resonance region}

\author{Tom Vrancx}
\email{Tom.Vrancx@UGent.be}
\author{Jan Ryckebusch}
\email{Jan.Ryckebusch@UGent.be}
\author{Jannes Nys}
\email{Jannes.Nys@UGent.be}

\affiliation{Department of Physics and Astronomy,\\
Ghent University, Proeftuinstraat 86, B-9000 Gent, Belgium}
\date{\today}

%######################################################################
\begin{abstract}
\begin{description}
\item[Background]
In $\pi^+n$ and $\pi^-p$ electroproduction, conventional models cannot satisfactory explain the data above the resonance region, in particular the transverse cross section. Although no high-energy \textsc{l}-\textsc{t}-separated cross-section data is available to date, a similar scenario can be inferred for $K^+\Lambda$ electroproduction.

\item[Purpose]
Develop a phenomenological model for the $p(\gamma^*,K^+)\Lambda$ reaction at forward angles and high-energies. Propose a universal framework for interpreting charged-kaon and charged-pion electroproduction above the resonance region.

\item[Method]
Guided by the recent model for charged-pion electroproduction, developed by the authors, a framework for $K^+\Lambda$ electroproduction at high energies and forward angles is constructed. To this end, a Reggeized background model for $K^+\Lambda$ photoproduction is first developed. This model is used as a starting base to set up an electroproduction framework.

\item[Results]
The few available data of the unseparated $p(\gamma^*,K^+)\Lambda$ cross section are well explained by the model. Predictions for the \textsc{l}-\textsc{t}-separation experiment planned with the 12 GeV upgrade at Jefferson Lab are given. The newly-proposed framework predicts an increased magnitude for the transverse structure function, similar to the situation in charged-pion electroproduction.

\item[Conclusions] 
Within a hadronic framework featuring Reggeized background amplitudes, $s$-channel resonance-parton effects can explain the observed magnitude of the unseparated $p(\gamma^*,K^+)\Lambda$ cross section at high energies and forward angles. Thereby, no hardening of the kaon electromagnetic form factor is required.

\end{description}
\end{abstract}

% 13.40.Gp  Electromagnetic form factors
% 13.60.Le  Meson production
% 24.10.-i  Nuclear reaction models and methods
% 25.30.Rw  Nuclear electroproduction reactions

\pacs{13.40.Gp, 13.60.Le, 24.10.-i, 25.30.Rw}

\maketitle

%######################################################################
\section{Introduction}
\label{sec:intro}
Above the resonance region, the transverse cross section $\s{t}$ measured in charged-pion electroproduction is significantly larger than predicted by regular hadronic models \cite{Blok:2008jy}. In Ref.\ \cite{Kaskulov:2010kf}, Kaskulov and Mosel proposed a framework explaining this observation. In the Kaskulov-Mosel formalism, the missing transverse strength is provided by the residual effects of nucleon resonances in the gauge-fixing $s$ (or $u$) channel. 

It is argued that such contributions become more important for increasing intermediate-proton and photon virtualities. Indeed, above the resonance region the proton is highly off-shell and the contributions from more massive intermediate states increase in importance. With growing intermediate-proton virtuality, also the hardening of the resonance electromagnetic transition can be anticipated to play an increasingly important role. This results in a dual viewpoint in which the residual effects can be interpreted as originating from the partonic picture of hadrons.

The resonance-parton (R-P) contributions are effectively implemented by means of an electromagnetic (EM) transition form factor for the proton in the $s$ channel. In Ref.\ \cite{Vrancx2013}, a new version of this form factor was proposed which has a simple physical interpretation and respects the correct on-shell limit. The resulting model was dubbed the ``Vrancx-Ryckebusch'' (VR) model and offers an explanation for the high-energy, forward-angle $\pi^+n$ and $\pi^-p$ electroproduction data, thereby covering a wide range of invariant masses ($2\GeV \lesssim W \lesssim 4\GeV$) and photon virtualities ($0.2\GeV^2 \lesssim Q^2 \lesssim 5\GeV^2$).

From the observations in the pion case, along with SU(3) symmetry considerations, one may infer that an increased transverse response might also occur in charged-kaon electroproduction. Within the VR framework, this can be anticipated from the employed strategy of introducing an effective EM transition form factor for the proton in the $s$ channel, accounting for the R-P contributions.

To this day, no $\s{t}$ data is available for high-energy $K^+\Lambda$ electroproduction and it is to be awaited if its magnitude is larger than expected. In this regard it is worth noting that the measured $p(\gamma^*,K^+)\Lambda$ unseparated cross section $\s{u}$ at high energies can be reproduced by the Vanderhaeghen-Guidal-Laget (VGL) model \cite{Guidal:1997hy, Vanderhaeghen:1997ts, Guidal:1999qi, Guidal:2003qs} after introducing an effective kaon EM form factor. The kaon EM cutoff energy employed in the VGL model is significantly increased compared to the value extracted from elastic $eK$ scattering. This may hint at an anomalously large transverse contribution to the unseparated cross section. After completing the 12 GeV upgrade at Jefferson Lab (JLab), one plans to measure the first $p(\gamma^*,K^+)\Lambda$ separated structure functions $\s{l}$ and $\s{t}$ at high energies \cite{Horn:2008pr}.

Following the strategy employed in charged-pion electroproduction \cite{Vrancx2013}, the VR model for high-energy forward-angle $K^+\Lambda$ electroproduction will be developed. In Sec.\ \ref{sec:model}, the transition currents are discussed. These will be used in Sec.\ \ref{sec:photoproduction} to construct an improved model for high-energy, forward-angle $K^+\Lambda$ photoproduction. Starting from this photoproduction model, the VR model will be derived in Sec.\ \ref{sec:electroproduction}. There, predictions will be presented for the above-mentioned experiment planned at JLab \cite{Horn:2008pr}. In Sec.\ \ref{sec:conclusions}, the conclusions of this work will be given.

%######################################################################
\section{Transition currents}
\label{sec:model}

%\subsection{Transition currents}
%\label{subsec:currents}

In complete analogy to the pion case \cite{Vrancx2013}, the adopted current for the gauged pseudoscalar-kaon exchange in $p(\gamma^*,K^+)\Lambda$ is given by
\begin{multline}
(J_K)^\mu_{\lambda_p,\lambda_\Lambda}(s, t, Q^2) = \\
i g_{K\Lambda p}\overline{u}_{\lambda_\Lambda}(p')\gamma_5\Biggl(F_{\gamma K K}(Q^2)\P_K(t, s)(2k' - q)^\mu\\
+ F_p(Q^2,s)\P'_K(t, s, Q^2)\frac{t - m_K^2}{s - m_p^2}(\feynp{p} + \feynq{q} + m_p)\gamma^\mu\Biggr)u_{\lambda_p}(p).\label{eq:J_K}
\end{multline}
Here, $p$, $q$, $k'$, and $p'$ are the four-momenta of the nucleon, of the virtual photon, of the kaon, and of the hyperon in the center-of-mass frame.\footnote{In Ref.\ \cite{Vrancx2013}, these four-momenta are also defined in the center-of-mass frame and not in the laboratory frame, as mentioned.} The Mandelstam variables $(s,t,u)$ are defined in the standard way. The proton and $\Lambda$ polarizations are denoted by $\lambda_p$ and $\lambda_\Lambda$, and the strong $K\Lambda p$ coupling constant by $g_{K\Lambda p}$. Note the absence of the SU(2) $\sqrt{2}$ factor in $J_K$, compared to the pion-exchange current of Ref.\ \cite{Vrancx2013}. The employed expression for the kaon-Regge propagator $\P_K(t, s)$ reads
\begin{align}
\P_K(t, s) = -\alpha_K'\varphi(\alpha_K(t))\Gamma(-\alpha_K(t))\biggl(\frac{s}{s_0}\biggr)^{\alpha_K(t)}.\label{eq:P_K}
\end{align}
Following the convention of the VGL \cite{Guidal:1997hy, Vanderhaeghen:1997ts, Guidal:1999qi, Guidal:2003qs} and the Regge-plus-Resonance (RPR) \cite{Corthals:2005ce, Corthals:2007kc, DeCruz:2011xi, DeCruz:2012bv} models, the ``mass scale'' $s_0$ is fixed to $s_0 = 1 \GeV^2$. In Ref.\ \cite{Vrancx2013}, the convention $s_0 = 1/\alpha'$ is adopted. The model assumptions with regard to the Regge trajectories $\alpha(t)$ and Regge phases $\varphi(\alpha(t))$, will be discussed in Sec.~\ref{subsec:third_trajectory}. The EM form factors $F_{\gamma K K}(Q^2)$ and $F_p(Q^2,s)$, and the modified kaon-Regge propagator $\P'_K(t, s, Q^2)$ will be discussed in Sec.\ \ref{subsec:form_factors}. At this point, it suffices to note that at vanishing photon virtuality $(Q^2=0)$ it holds that $F_{\gamma K K}(Q^2=0) = F_p(Q^2=0,s) = 1$ and $\P'_K(t, s, Q^2=0) \equiv \P_K(t, s)$.

The expressions for the vector ($V$, $J^P = 1^-$) and axial-vector ($A$, $J^P = 1^+$) transition currents $J_{K_V}$ and $J_{K_A}$ are adopted from Ref.\ \cite{Kaskulov:2010kf} and read
\begin{multline}
(J_{K_V})^\mu_{\lambda_p,\lambda_\Lambda}(s, t, Q^2) = \\
G_{\gamma K_V K} G_{K_V \Lambda p} F_{\gamma K_V K}(Q^2) \P_{K_V}(t, s) \varepsilon^{\mu\nu\sigma\tau} q_\nu k_\sigma \\
\times \overline{u}_{\lambda_\Lambda}(p')\Biggl((1 + \kappa_{K_V \Lambda p})\gamma_\tau - \frac{\kappa_{K_V \Lambda p}}{2m_p}(p + p')_\tau \Biggr)u_{\lambda_p}(p),
\label{eq:J_KV}
\end{multline}
and
\begin{multline}
(J_{K_A})^\mu_{\lambda_p,\lambda_\Lambda}(s, t, Q^2) = \\
-i G_{\gamma K_A K} G_{K_A \Lambda p} F_{\gamma K_A K}(Q^2) \P_{K_A}(t, s) (k^\mu q^\nu - q^\sigma k_\sigma g^{\mu\nu}) \\
\times \overline{u}_{\lambda_\Lambda}(p')\Biggl((1 + \kappa_{K_A \Lambda p})\gamma_\nu - \frac{\kappa_{K_A \Lambda p}}{2m_p}(p + p')_\nu \Biggr)\gamma_5u_{\lambda_p}(p).
\label{eq:J_A}
\end{multline}
Note that the $\sqrt{2}$ factor was dropped again, and that a minus sign missing in Ref.\ \cite{Kaskulov:2010kf}, was added to the right-hand side of Eq.\ \eq{eq:J_A}. The EM coupling constant is represented by $G_{\gamma K_{V,A} K}$, and the standard vector and anomalous tensor coupling constants by $G_{K_{V,A} \Lambda p}$ and $\kappa_{K_{V,A}\Lambda p}$. The vector and axial-vector Regge propagators have the same functional dependence and are given by
\begin{align}
\P_{K_{V,A}}(t, s) = -\alpha_{K_{V,A}}'\varphi(\alpha_{K_{V,A}}(t) - 1)\Gamma(1 - \alpha_{K_{V,A}}(t))\biggl(\frac{s}{s_0}\biggr)^{\alpha_{K_{V,A}}(t) - 1}.
\end{align}
The EM transition form factors $F_{\gamma K_{V,A} K}(Q^2)$ will be discussed in Sec.\ \ref{subsec:form_factors}. As for the form factors $F_{\gamma K K}(Q^2)$ and $F_p(Q^2,s)$ in Eq.\ \eq{eq:J_K}, it holds that $F_{\gamma K_{V,A} K}(0) = 1$.

%\subsection{Regge trajectories}
%\label{subsec:kaon_regge}

%######################################################################
\section{High-energy \texorpdfstring{$K^+\Lambda$}{K+ Lambda} photoproduction}
\label{sec:photoproduction}

\subsection{A third Regge trajectory}
\label{subsec:third_trajectory}
In the kaon sector, the two most important Regge trajectories are the $K(494)$ (pseudoscalar) and $K^*(892)$ (vector) trajectories \cite{Beringer:1900zz}. These can be parametrized as \cite{Corthals:2005ce}
\begin{align}
\alpha_K(t) &= \alpha'_K(t - m_K^2),\\
\alpha_{K^*}(t) &= 1 + \alpha'_{K^*}(t - m_{K^*}^2),
\end{align}
with $\alpha'_K = 0.70\GeV^{-2}$ and $\alpha'_{K^*} = 0.85\GeV^{-2}$. Both the VGL and the RPR model feature these two trajectories and have established that they are essential for the description of $K^+\Lambda$ photo- and electroproduction.

\begin{table*}[!htbp]
\caption{Coupling constants and corresponding $\cndf$ values of the three-trajectory Regge model variants featuring a $K$, a $K^*$ and a $K_1(1400)$ trajectory. Results are listed for a rotating $K$ trajectory and all phase combinations for the $K^{*}$ and $K_{1}$. For the sake of reference also the results of the two-trajectory Regge-2011 model are shown. The models were optimized against the high energy ($W > 2.6\GeV$) and forward-angle ($\cos\thcm > 0.35$) CLAS data for the $p(\gamma,K^+)\Lambda$ differential cross section and recoil polarization (262 data points) \cite{McCracken:2009ra}.}
\label{tab:chi2-K1_1400}
\renewcommand{\arraystretch}{1.15}
\begin{tabular*}{\textwidth}{@{\extracolsep{\fill}}cc*{1}......}
\hline
\noalign{\smallskip}\hline
Model & $\{\varphi_K, \varphi_{K^*}, \varphi_{K_1(1400)}\}$ & \multicolumn{1}{c}{$g_{K \Lambda p}$} & \multicolumn{1}{c}{$G_{\gamma K^*K} G_{K^* \Lambda p}$ $(\textrm{GeV}^{-1})$} & \multicolumn{1}{c}{$\kappa_{K^*\Lambda p}$} & \multicolumn{1}{c}{$G_{\gamma K_1(1400)K} G_{K_1(1400) \Lambda p}$ $(\textrm{GeV}^{-1})$} & \multicolumn{1}{c}{$\kappa_{K_1(1400)\Lambda p}$} & \multicolumn{1}{c}{$\cndf$}\\
\hline
% CCC 34.6
% CRC 35.2
% CCR 34.9
% CRR 32.5
Ia & $\{R,C,C\}$ & -12.2 & -0.29 & 50 & 32.3 & -1.35 & 17.1 \\
IIa & $\{R,R,C\}$ & -13.2 & -9.84 & 1.51 & 16.5 & -0.86 & 2.58 \\
IIIa & $\{R,C,R\}$ & -10.8 & -0.44 & 50 & 28.0 & -1.74 & 13.6 \\
IVa & $\{R,R,R\}$ & -12.6 & -10.5 & 1.38 & 7.48 & -0.19 & 2.99 \\
Regge-2011 & $\{R,R,-\}$ & -12.9 & -10.8 & 1.77 & - & - & 3.15 \\
\hline
\noalign{\smallskip}\hline
\end{tabular*}

\end{table*}
\begin{table*}[!htbp]
\caption{As in Table \ref{tab:chi2-K1_1400}, but for the model based on a third $K^*(1410)$ trajectory instead of a $K_1(1400)$ trajectory.}
\label{tab:chi2-K*1410}
\renewcommand{\arraystretch}{1.15}
\begin{tabular*}{\textwidth}{@{\extracolsep{\fill}}cc*{1}......}
\hline
\noalign{\smallskip}\hline
Model & $\{\varphi_K, \varphi_{K^*}, \varphi_{K^*(1410)}\}$ & \multicolumn{1}{c}{$g_{K \Lambda p}$} & \multicolumn{1}{c}{$G_{\gamma K^*K} G_{K^* \Lambda p}$ $(\textrm{GeV}^{-1})$} & \multicolumn{1}{c}{$\kappa_{K^*\Lambda p}$} & \multicolumn{1}{c}{$G_{\gamma K^*(1410)K} G_{K^*(1410) \Lambda p}$ $(\textrm{GeV}^{-1})$} & \multicolumn{1}{c}{$\kappa_{K^*(1410)\Lambda p}$} & \multicolumn{1}{c}{$\cndf$}\\
\hline
% CCC 34.1
% CRC 26.3
% CCR 26.3
% CRR 17.2
Ib & $\{R,C,C\}$ & -13.2 & -0.01 & 0.03 & 41.6 & -0.53 & 6.33 \\
IIb & $\{R,R,C\}$ & -13.2 & -6.79 & 1.04 & 32.7 & 0.70 & 1.06 \\
IIIb & $\{R,C,R\}$ & -12.4 & -0.23 & 50 & 49.3 & -0.46 & 4.91 \\
IVb & $\{R,R,R\}$ & -14.2 & -19.4 & 0.68 & -54.4 & -0.55 & 2.04 \\
Regge-2011 & $\{R,R,-\}$ & -12.9 & -10.8 & 1.77 & - & - & 3.15 \\
\hline
\noalign{\smallskip}\hline
\end{tabular*}
\end{table*}

In Refs.~\cite{DeCruz:2011xi, DeCruz:2012bv}, the Regge background for the $p(\gamma,K^+)\Lambda$ reaction was determined from the recent differential cross section and recoil polarization data by the CEBAF Large Acceptance Spectrometer (CLAS) Collaboration \cite{McCracken:2009ra}.\footnote{In Table IV of Ref.\ \cite{DeCruz:2012bv}, an overview is given of the available $p(\gamma,K^+)\Lambda$ data.} More specifically, a Bayesian analysis was performed for the high-energy ($W > 2.6\GeV$) and forward-angle ($\cos\thcm > 0.35$) part of these CLAS data (262 data points) to determine the Regge model variant with the highest evidence. Here, $W = \sqrt{s}$ is the invariant mass and $\thcm$ the kaon scattering angle in the center-of-mass frame. It was found that the best model, dubbed ``Regge-2011'', features rotating phases for both the $K$ and $K^*$ trajectories. For this model, a $\cndf = 3.15$ is obtained for the description of the high-energy and forward-angle $p(\gamma,K^+)\Lambda$ CLAS data \cite{DeCruz:2011xi, DeCruz:2012bv}. As there is definitely room for improvement, the possibility is exploited of introducing a third Regge trajectory contributing to the $p(\gamma^{(*)}, K^+)\Lambda$ reaction. In this regard, two candidates are considered: the $K_1(1400)$ (axial-vector) and the $K^*(1410)$ (vector) trajectory \cite{Beringer:1900zz}. These are parametrized as \cite{Corthals:2006nz, VancraeyveldPhD}
\begin{align}
\alpha_{K_1(1400)}(t) &= 1 + \alpha'_{K_1(1400)}(t - m_{K_1(1400)}^2),\\
\alpha_{K^*(1410)}(t) &= 1 + \alpha'_{K^*(1410)}(t - m_{K^*(1410)}^2),
\end{align}
with $\alpha'_{K_1(1400)} = 0.75\GeV^{-2}$ and $\alpha'_{K^*(1410)} = 0.83 \GeV^{-2}$. All the trajectories considered here are degenerate. This means that the corresponding Regge phases can either be constant ($C$) or be rotating ($R$):
\begin{align}
\varphi(\alpha(t)) =
\begin{cases}
1 &C,\\
e^{-i\pi\alpha(t)} &R.
\end{cases}
\end{align}

\subsection{Parameter constraints}
\label{subsec:model_restrictions}
Since the phases of the Regge trajectories considered here can be either constant or rotating, there are 8 possible variants for each three-trajectory model. These models, however, are all restricted to some extent as the $K$ and $K^*$ coupling constants must meet certain constraints, based on symmetry arguments. The strong coupling $g_{K \Lambda p}$ can be inferred from the strong pion-nucleon coupling $g_{\pi N N}$ by means of SU(3) symmetry:
\begin{align}
g_{K \Lambda p} = -\frac{1}{\sqrt{3}}(3 - 2\alpha_D)g_{\pi N N},
\end{align}
with $\alpha_D = 0.644$ the experimentally determined SU(3) symmetric coupling fraction. By allowing a 20\% breaking of SU(3) symmetry and taking into account the uncertainty on the pion-nucleon coupling, i.e.\ $g_{\pi N N} \simeq \hbox{13.0--13.5}$ \cite{Vrancx2013}, the following limits on $g_{K \Lambda p}$ emerge:
\begin{align}
-16.0 \le g_{K \Lambda p} \le -10.3.\label{eq:constraint_1}
\end{align}
The EM coupling constant $G_{\gamma K^*K}$ can be estimated from the decay width of $K^* \to K\gamma$ \cite{Guidal:1997hy}:
\begin{align}
\Gamma_{K^* \to K\gamma} = \frac{\alpha_e}{24}\frac{G_{\gamma K^*K}^2}{m_{K^*}^3}(m_{K^*}^2 - m_K^2)^3,
\end{align}
with $\alpha_e$ the fine-structure constant. From the experimentally determined value $\Gamma_{K^* \to K\gamma} = 50 \pm 5\keV$ \cite{Beringer:1900zz} one obtains 
\begin{align}
G_{\gamma K^*K} = 0.834 \pm 0.042 \GeV^{-1}.
\end{align}

\begin{figure}[!b]
\centering
\includegraphics{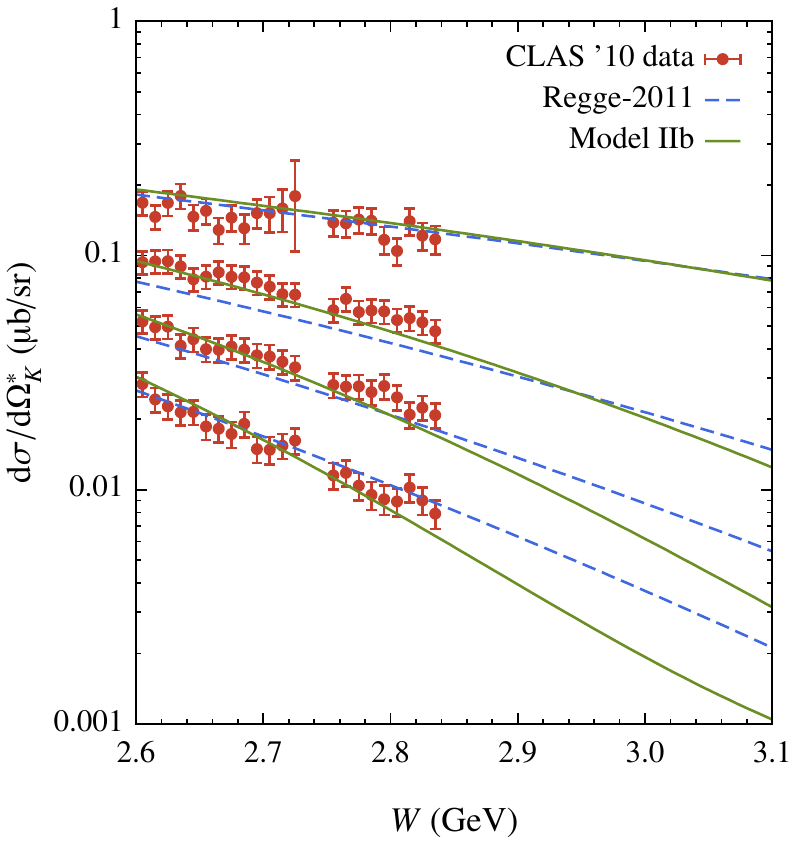}
\caption{(Color online) The $W$ dependence of the $p(\gamma,K^+)\Lambda$ unpolarized differential cross section for (from bottom to top) $\cos\thcm = 0.50, \,0.60, \,0.70, \,\hbox{0.86--0.87}$. Predictions from the Regge-2011 model and model IIb are shown. The data are from Ref.\ \cite{McCracken:2009ra}.}
\label{fig:photo_CLAS-diffcs}
\end{figure}

Also the strong vector and tensor couplings $G_{K^* \Lambda p}$ and $\kappa_{K^* \Lambda p}$ can be related to $G_{\rho N N}$ and $\kappa_{\rho N N}$ through SU(3) symmetry. However, following the arguments given in Ref.\ \cite{Guidal:1997hy}, only the predicted signs for the vector and tensor couplings will be respected:
\begin{align}
G_{K^* \Lambda p} < 0, \hspace{0.1\columnwidth}
\kappa_{K^* \Lambda p} > 0.\label{eq:constraint_2}
\end{align}
Due to the lack of relevant experimental information, no constraints are imposed on the $K_1(1400)$ and $K^*(1410)$ coupling constants.

\subsection{Results}
\label{subsec:results_photo}

Tables \ref{tab:chi2-K1_1400} and \ref{tab:chi2-K*1410} list the best-fit parameters of the three-trajectory model variants. The coupling constants are optimized against the high-energy and forward-angle CLAS data and respect the constraints of Eqs.~\eq{eq:constraint_1} and \eq{eq:constraint_2}. Only the models featuring a rotating $K$ trajectory are listed, as those with a constant $K$ trajectory are not compatible with the data. Indeed, a constant phase for the $K$ trajectory leads to $\cndf = \hbox{32.5--35.2}$ for the $K_1(1400)$ model variants, and to $\cndf = \hbox{17.2--34.1}$ for the $K^*(1410)$ model variants. 

\begin{figure}[!t]
\centering
\includegraphics{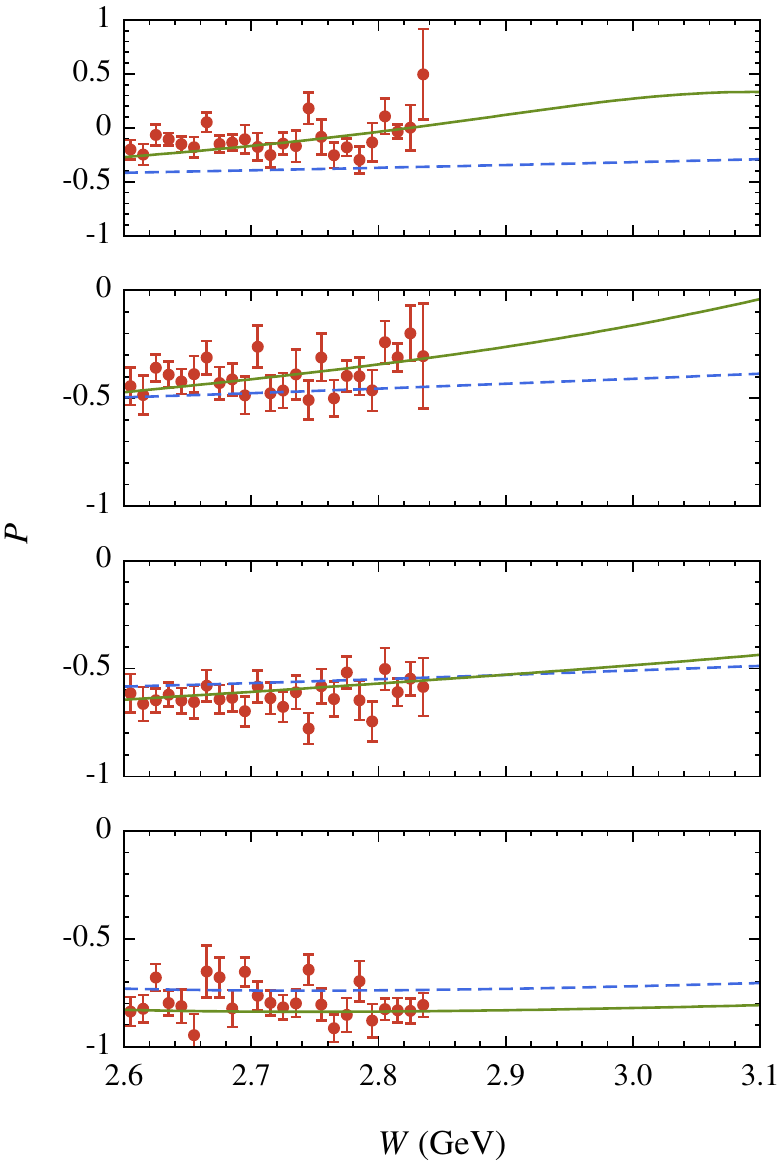}
\caption{(Color online) The $W$ dependence of the $p(\gamma,K^+)\Lambda$ recoil polarization $P$ for (from bottom to top) $\cos\thcm = 0.50, \,0.60, \,0.70, \,\hbox{0.86--0.87}$. Predictions from the Regge-2011 model and model IIb are shown. Curve notations of Fig.\ \ref{fig:photo_CLAS-diffcs} are used. The data are from Ref.\ \cite{McCracken:2009ra}.}
\label{fig:photo_CLAS-rec}
\end{figure}

The models with a rotating $K$ and constant $K^*$ phase are systematically in poorest agreement with the data. In fact, these models yield coupling constants approaching the maximum values allowed during the optimization process: models Ia, IIIa, and IIIb yield $\kappa_{K^*\Lambda p} = 50$, and model Ib yields $G_{\gamma K^*K} G_{K^* \Lambda p} = -0.01\allowbreak\GeV^{-1}$. This implies that the analyzed CLAS data exclude a constant $K^*$ phase given the constraints of Eq.\ \eq{eq:constraint_2}.

Amongst the models with a rotating $K^*$ phase, those with a $K^*(1410)$ (vector) trajectory perform better than those with an $K_1(1400)$ (axial-vector) trajectory. Model IIb clearly stands out from the rest and is in excellent agreement with the data ($\cndf = 1.06$). This model features a constant $K^*(1410)$ and rotating $K$ and $K^*$ trajectories. Note that the value $g_{K \Lambda p} = -13.2$ for this model coincides with the predicted SU(3) value, given the uncertainty on $g_{\pi N N}$. This is also the case for the Regge-2011 model. The employed CLAS data, along with the corresponding predictions of the Regge-2011 model and model IIb, are shown in Figs.\ \ref{fig:photo_CLAS-diffcs} and \ref{fig:photo_CLAS-rec} for four $\cos\thcm$ bins. Model IIb constitutes the basis for the VR model, which will be discussed in the forthcoming section.

\begin{figure}[!b]
\centering
\includegraphics{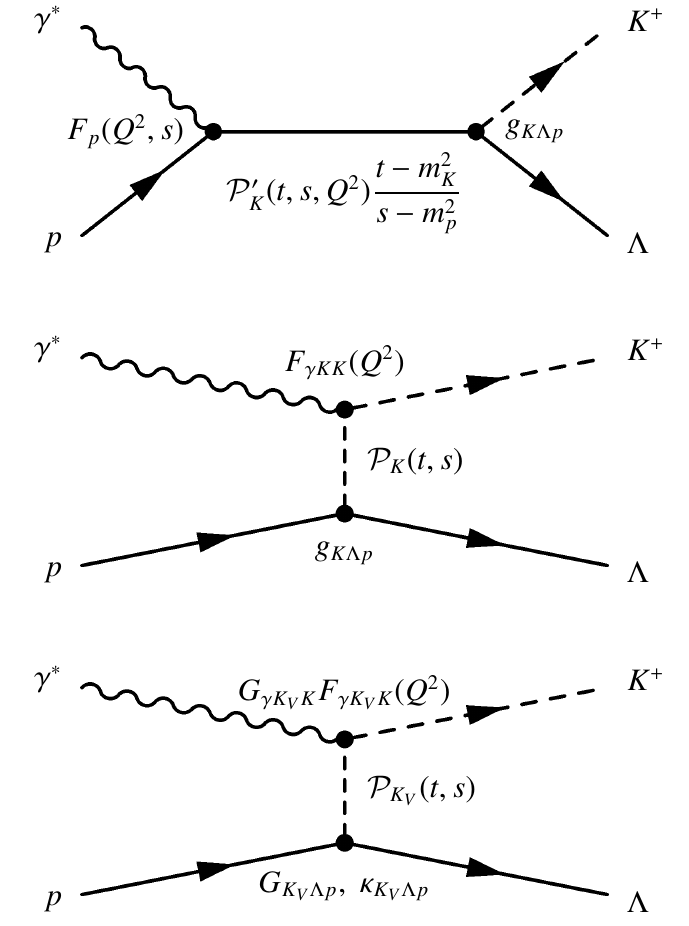}
\caption{The gauge-fixing $s$-channel diagram and the Reggeized pseudoscalar and vector $t$-channel diagrams that constitute the VR model for $K^+\Lambda$ electroproduction above the resonance region.}
\label{fig:scheme_VR}
\end{figure}

%######################################################################
\section{High-energy \texorpdfstring{$K^+\Lambda$}{K+ Lambda} electroproduction}
\label{sec:electroproduction}

\subsection{Form factors}
\label{subsec:form_factors}

As the $Q^2=0$ limit of the proposed $p(\gamma^*,K^+)\Lambda$ model has been established, the $Q^2$-dependent quantities in the transition current operators of Eqs.\ \eq{eq:J_K} and \eq{eq:J_KV} can now be examined. Pursuing the analogy to the VR model for pion electroproduction, an antishrinkage effect is introduced in the $s$-channel gauge-fixing term of the kaon transition current \eq{eq:J_K}. To this end, the Regge propagator $\P'_K(t, s, Q^2)$ in Eq.\ \eq{eq:J_K} is defined as in Eq.\ \eq{eq:P_K}, but with an altered Regge slope:
\begin{align}
\alpha_K' \to \alpha_K'(Q^2, s) = \frac{\alpha_K'}{1 + a\frac{Q^2}{s}}.\label{eq:anti-shrinkage}
\end{align}
Here, $a$ is the corresponding dimensionless slope parameter, which has yet to be determined. Figure \ref{fig:scheme_VR} depicts the $s$- and $t$-channel diagrams which constitute the VR model for $p(\gamma^*,K^+)\Lambda$.

A monopole form is adopted for the elastic kaon EM form factor $F_{\gamma K K}(Q^2)$ in of Eq.~\eq{eq:J_K} with a kaon cutoff energy $\Lambda_{\gamma K K}$:
\begin{align}
F_{\gamma K K}(Q^2) = \Biggl(1 + \frac{Q^2}{\Lambda_{\gamma K K}^2}\Biggr)^{-1}.\label{eq:F_gKK}
\end{align}
As the root-mean-square charge radius of the $K$ is experimentally determined as \cite{Beringer:1900zz}
\begin{align}
\sqrt{\langle r_K^2 \rangle} = 0.560 \pm 0.031 \fm,
\end{align}
the corresponding monopole cutoff energy is 
\begin{align}
\Lambda_{\gamma K K} = \sqrt{\frac{6}{\langle r_K^2 \rangle}} = 863 \pm 48 \MeV.\label{eq:cutoff_K}
\end{align}
In the vector-meson dominance (VMD) model, the kaon EM form factor receives contributions from primarily the $\rho$, $\omega$, and $\phi$ mesons \cite{Ivashyn:2006gf}:
\begin{align}
F_{\gamma K K}^{\textrm{VMD}}(Q^2) = \frac{1}{N}\sum_{v = \rho, \omega, \phi}\frac{g_{vKK}}{f_v}\frac{1}{1 + Q^2/m_v^2},\label{eq:F_VMD}
\end{align}
with $N = \sum_{v = \rho, \omega, \phi}\frac{g_{vKK}}{f_v}$ a normalization constant. Assuming an exact SU(3) flavor symmetry, the EM and strong $\omega$ and $\phi$ coupling constants can be related to those of the $\rho$: 
\begin{align}
f_\omega &= 3f_\rho, &g_{\omega K K} &= g_{\rho K K},\nonumber\\
f_\phi &= -\frac{3}{\sqrt{2}}f_\rho, &g_{\phi K K} &= -\sqrt{2}g_{\rho K K}.\label{eq:SU3_symmetry}
\end{align}
From these SU(3) coefficients and the masses of the $\rho$, $\omega$, and $\phi$ mesons \cite{Beringer:1900zz}, the VMD monopole cutoff energy for the $K$ is calculated as
\begin{align}
\Lambda_{\gamma K K}^{\textrm{VMD}} = \left(-\frac{\d F_{\gamma K K}^{\textrm{VMD}}(Q^2)}{\d Q^2}\Biggr|_{Q^2 = 0}\right)^{-1/2} \simeq 838\MeV,\label{eq:cutoff_K-VMD}
\end{align}
which is consistent with the experimental value of Eq.\ \eq{eq:cutoff_K}.

The form factors $F_{\gamma K_V K}(Q^2)$ in Eq.~\eq{eq:J_KV} describe the EM transitions of the vector-kaon trajectories to the outgoing pseudoscalar kaon. For these form factors a monopole form \eq{eq:F_gKK} is also adopted. No data is available for the cutoff energies $\Lambda_{\gamma K^* K}$ and $\Lambda_{\gamma K^*(1410) K}$, however, so one has to rely on the corresponding VMD predictions. The VMD description requires the following replacement in expression \eq{eq:F_VMD}:
\begin{align}
g_{vKK} \to G_{v K_V K}.
\end{align}
As the $K_V$ are nothing but orbitally excited states of the $K$, the same SU(3) constraints \eq{eq:SU3_symmetry} apply to the strong coupling constants $G_{V K_V K}$:
\begin{align}
G_{\omega K_V K} &= G_{\rho K_V K}, \nonumber\\
G_{\phi K_V K} &= -\sqrt{2}G_{\rho K_V K}.
\end{align}
Therefore, the $K^*$ and $K^*(1410)$ EM transition form factors in the VMD model are identical and equal to $F_{\gamma K K}^{\textrm{VMD}}(Q^2)$. Consequently, the value of Eq.\ \eq{eq:cutoff_K-VMD} will be used for the corresponding cutoff energies:
\begin{align}
\Lambda_{\gamma K_V K} = 838\MeV.
\end{align}
Note that the above reasoning also applies to axial-vector kaons.

The form factor $F_p(Q^2, s)$ in Eq.\ \eq{eq:J_K} describes the EM transition of an on-shell to an off-shell proton with squared four-momentum $s$, induced by a virtual photon. In the VR model for pion electroproduction, $F_p(Q^2, s)$ is a dipole \cite{Vrancx2013}:
\begin{align}
F_p(Q^2, s) = \Biggl(1 + \frac{Q^2}{\Lambda_{\gamma p p^*}^2(s)}\Biggr)^{-2},\label{eq:proton_EMTFF}
\end{align}
with an $s$-dependent cutoff energy ($s \ge m_p^2$)
\begin{align}
\Lambda_{\gamma p p^*}(s) = \Lambda_{\gamma pp} + (\Lambda_\infty - \Lambda_{\gamma pp})\Biggl(1 - \frac{m_p^2}{s}\Biggr).
\end{align}
Here, $\Lambda_{\gamma pp} = 840\MeV$ is the on-shell proton EM cutoff energy. The asymptotic, off-shell proton cutoff energy was determined as $\Lambda_\infty = 2194 \MeV$ \cite{Vrancx2013}.

\begin{figure}[!t]
\centering
\includegraphics{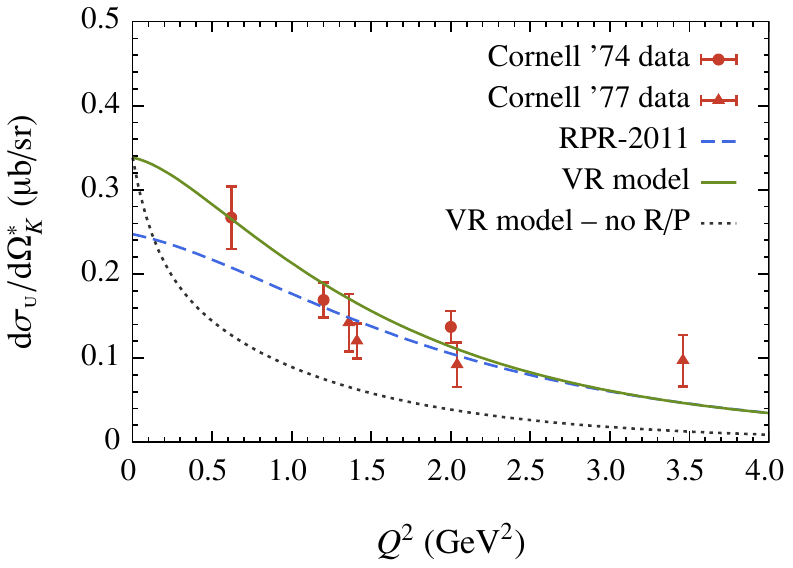}
\caption{(Color online) The $Q^2$ dependence of the $p(\gamma^*,K^+)\Lambda$ unseparated cross section $\d\s{u}/\d \Omega_K^*$. Predictions from the RPR model, the VR model, and the VR model without R-P contributions ($\Lambda_\infty = \Lambda_{\gamma p p}$ and $a = 0$) are shown for $W =\allowbreak 2.70\GeV$, $\cos\thcm = 0.98$, and $\varepsilon = 0.86$, which are the averaged kinematics for the different datasets \cite{Bebek:1974bt, Bebek:1976qg}.}
\label{fig:electro_Cornell}
\end{figure}

\begin{figure}[!Hb]
\centering
\includegraphics{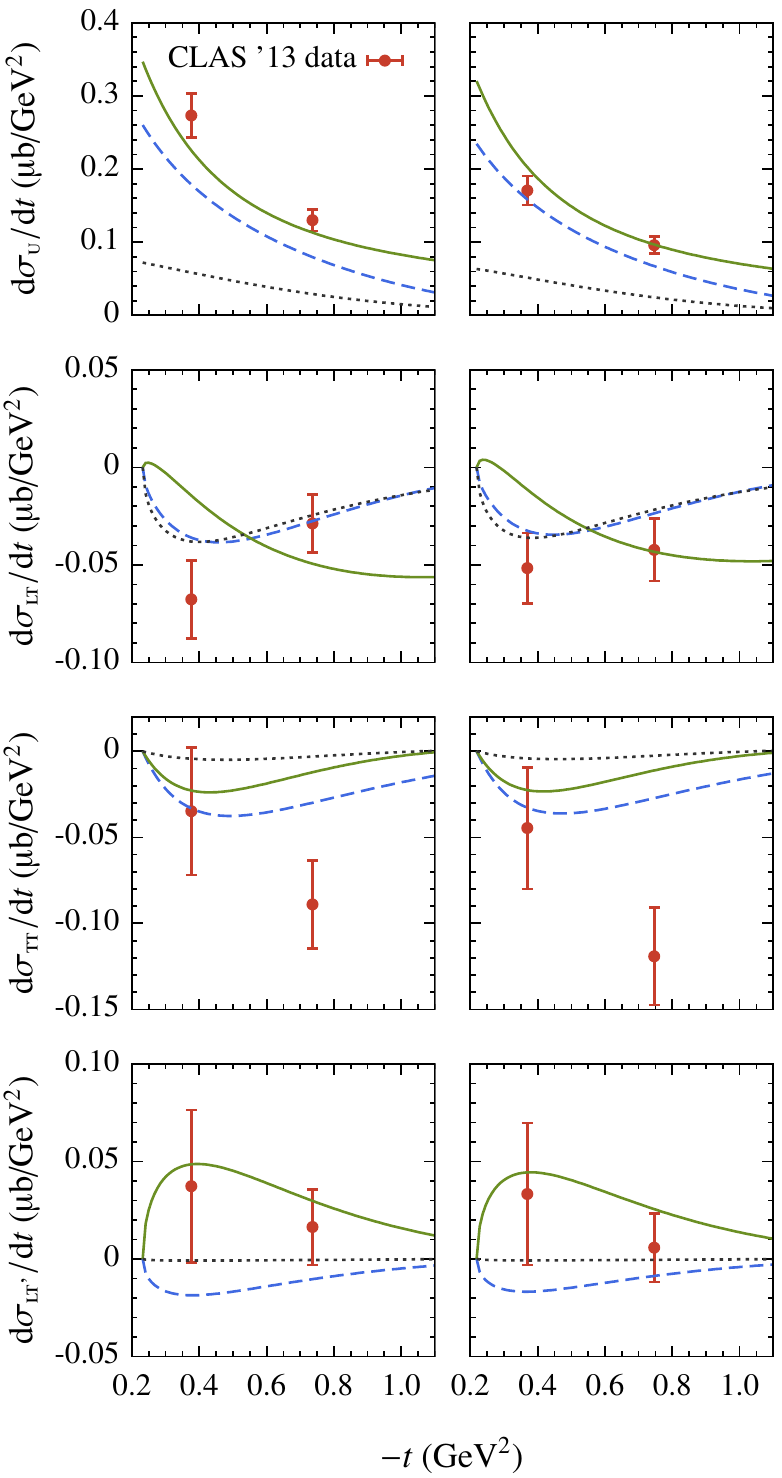}
\caption{(Color online) The $-t$ dependence of the $p(\gamma^*,K^+)\Lambda$ unseparated cross section $\d\s{u}/\d t$ and the separated cross sections $\d\s{lt}/\d t$, $\d\s{tt}/\d t$, and $\d\s{lt'}/\d t$ at $E_e = 5.499\GeV$ and $Q^2 = 1.80\GeV^2$ for $W = 2.525 \GeV$ (left) and $W = 2.575\GeV$ (right). Curve notations of Fig.~\ref{fig:electro_Cornell} are used. The data are from Ref.~\cite{Carman:2012qj}.}
\label{fig:electro_CLAS}
\end{figure}

\subsection{Results}
\label{subsec:results_electro}

The value of $a$ in Eq.~\eq{eq:anti-shrinkage} is the only parameter left in the VR model and is fitted to the scarce high-energy, forward-angle $p(\gamma^*,K^+)\Lambda$ data. In order to tune the VR model for pion electroproduction, data with $-t \lesssim 0.5\GeV^2$ was used \cite{Vrancx2013}. As few $p(\gamma^*,K^+)\Lambda$ data are available that cover the high-energy region, this range will be extended to $-t < 1 \GeV^2$. For the same reason the minimum $W$ value will be decreased from $2.6 \GeV$ (Sec.\ \ref{sec:photoproduction}) to $2.5\GeV$. There are 25 published data points that meet these kinematic restrictions: 9 data points measured at Cornell in the seventies \cite{Bebek:1974bt, Bebek:1976qg, Bebek:1977bv} and 16 recent data points from CLAS \cite{Carman:2012qj}. Most of the $p(\gamma^*,K^+)\Lambda$ data are available at $W < 2.5\GeV$, recent examples of which can be found in Refs.\ \cite{Mohring:2002tr, Ambrozewicz:2006zj, Nasseripour:2008aa, Carman:2009fi, Coman:2009jk}. For the 25 high-energy and forward-angle data points, the optimum value for the slope parameter is found to be
\begin{align}
a = 2.43.
\end{align}
Remarkably, this value coincides with the one obtained in the pion case \cite{Vrancx2013}. With $\cndf = 2.93$, the resulting VR model provides a fair description of the considered $p(\gamma^*,K^+)\Lambda$ data. A word of caution is in order, given the scarcity of the data and the fact that they cover a rather limited $W$ range.

Figures \ref{fig:electro_Cornell} and \ref{fig:electro_CLAS} show 23 of the 25 employed data points, along with the corresponding predictions of the VR and RPR-2011 models. The RPR-2011 model is a prototypical example of a single-channel model, designed to describe the $p(\gamma^{(*)},K^+)\Lambda$ reaction both in and beyond the resonance region \cite{DeCruz:2011xi, DeCruz:2012bv}. It yields $\cndf = 3.58$ for the high-energy $p(\gamma^*,K^+)\Lambda$ data considered. In addition to the $t$-channel Regge-2011 background discussed in Sec.~\ref{subsec:third_trajectory}, RPR-2011 features the exchange of the nucleon resonances $S_{11}(1535)$, $S_{11}(1650)$, $F_{15}(1680)$, $P_{13}(1720)$, $D_{13}(1900)$, $P_{13}(1900)$, $P_{11}(1900)$, and $F_{15}(2000)$, in the $s$ channel. For $W \gtrsim 2 \GeV$, the effects of these resonances is rather modest.

The VR model is in good agreement with the 19 $\s{u}$ data points. For $Q^2 \lesssim 2 \GeV^2$, the VR model predicts larger $\s{u}$ cross sections than Regge-2011. Both models predict a similar $\s{u}$ for $3 \lesssim Q^2 \lesssim 4\GeV^2$, but have different $Q^2\to\infty$ limits. Fig.~\ref{fig:electro_CLAS} contains the available data for the separated cross sections at $W > 2.5\GeV$. The biggest difference between the VR and RPR-2011 models is observed for the $\s{lt}$ and $\s{lt'}$. The largest deviations between theory and data are found for the $\s{lt}$ and $\s{tt}$. The quantity and quality of the data, however, does not allow one to draw any definite conclusions.

\begin{figure}[!Hb]
\centering
\includegraphics{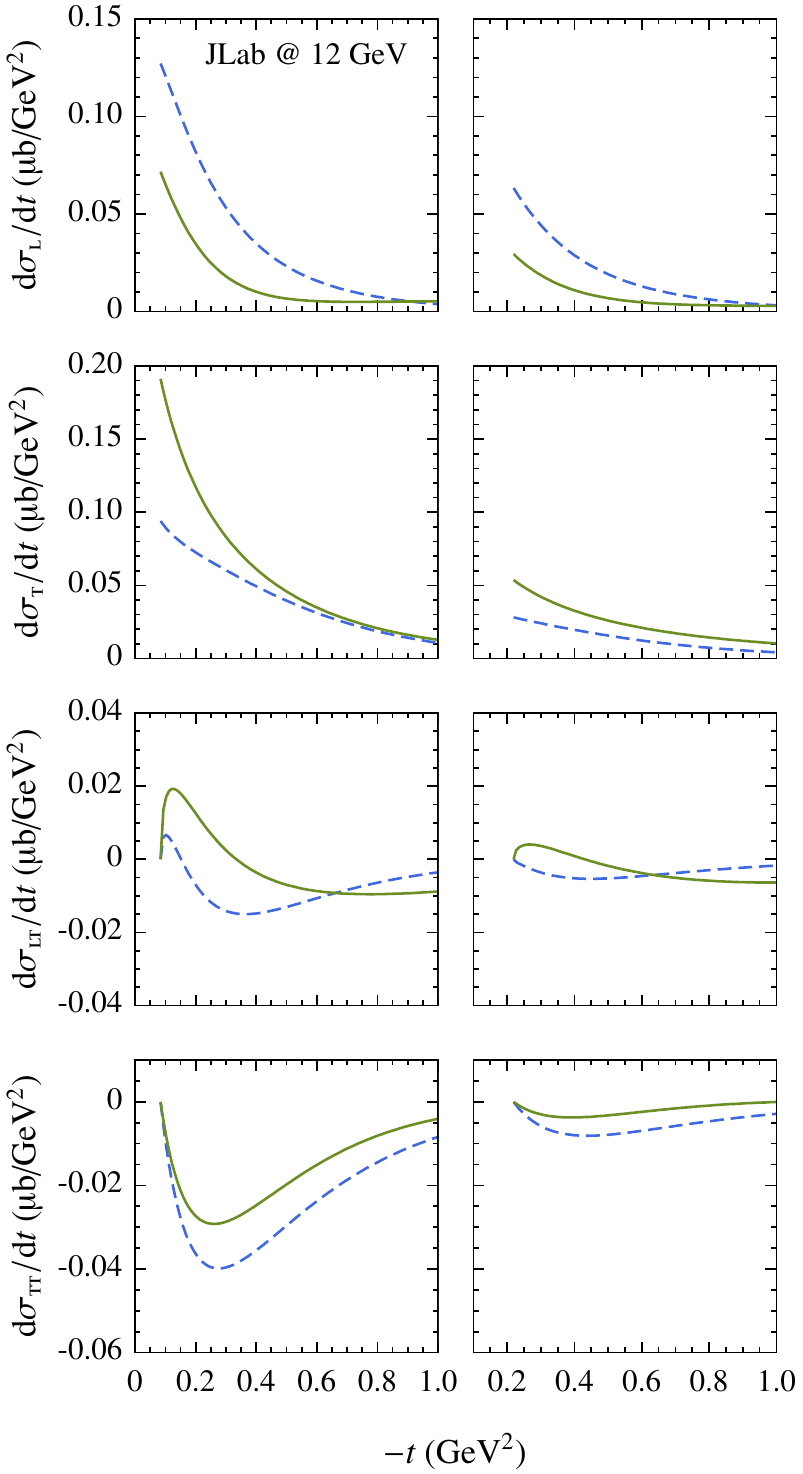}
\caption{(Color online) The $-t$ dependence of the separated $p(\gamma^*,K^+)\Lambda$ cross sections $\d\s{l}/\d t$, $\d\s{t}/\d t$, $\d\s{lt}/\d t$, and $\d\s{tt}/\d t$ at $W = 3.14\GeV$ for $Q^2 = 1.25 \GeV^2$ (left) and $Q^2 = 3.00\GeV^2$ (right). Curve notations of Fig.\ \ref{fig:electro_Cornell} are used. These are predictions for the planned $p(\gamma^*,K^+)\Lambda$ \textsc{l}-\textsc{t}-separation experiment \cite{Horn:2008pr}.}
\label{fig:electro_JLab12}
\end{figure}

\begin{figure}[!Ht]
\centering
\includegraphics{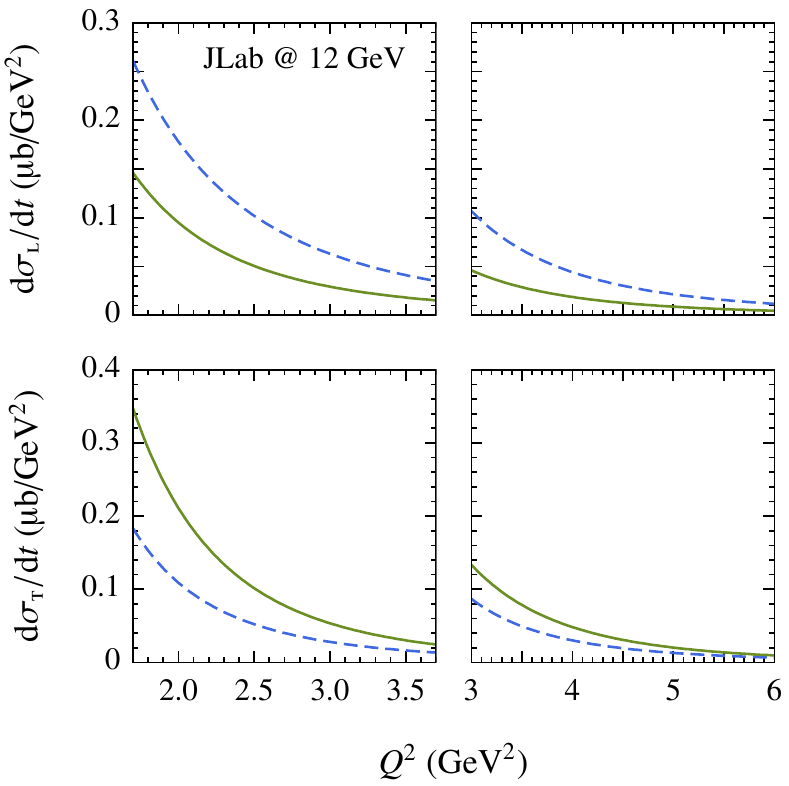}
\caption{(Color online) The $Q^2$ dependence of the separated $p(\gamma^*,K^+)\Lambda$ cross sections $\d\s{l}/\d t$, $\d\s{t}/\d t$, $\d\s{lt}/\d t$, and $\d\s{tt}/\d t$ at $t = t_\text{min}$ for $x_B = 0.25$ (left) and $x_B = 0.40$ (right), with $x_B$ being the Bjorken scaling variable. For the shown $Q^2$ ranges, one has $2.45\GeV \lesssim W \lesssim 3.46\GeV$ and $0.21\GeV^2 \lesssim -t \lesssim 0.25\GeV^2$ for $x_B = 0.25$, and $2.32\GeV \lesssim W \lesssim 3.14\GeV$ and $0.50\GeV^2 \lesssim -t \lesssim 0.53\GeV^2$ for $x_B = 0.40$. Curve notations of Fig.\ \ref{fig:electro_Cornell} are used. These are predictions for the planned $p(\gamma^*,K^+)\Lambda$ \textsc{l}-\textsc{t}-separation experiment of Ref.~\cite{Horn:2008pr}.}
\label{fig:electro_JLab12-Q2}
\end{figure}

From Figs.\ \ref{fig:electro_Cornell} and \ref{fig:electro_CLAS}, one can easily see that in the VR model the anomalously large $\s{u}$ can be attributed to the R-P effects. An appealing feature of this approach is that $F_p(Q^2, s)$ can account for both the pion \cite{Vrancx2013} and the kaon data at high energies and forward angles. It is worth mentioning that the RPR-2011 model does not adopt a proton EM transition form factor, i.e.\ $F^\text{RPR}_p(Q^2, s) = 1$. As a competing explanation for the observed trends in the $Q^2$ evolution of the data, a hard form factor is introduced at the $K$ and $K^*$ EM vertices of the RPR-2011 model:
\begin{align}
\Lambda_{\gamma K K}^\text{RPR} = \Lambda_{\gamma K^* K}^\text{RPR} = 1.3 \GeV.
\end{align}
For the $K$, this is a considerably larger cutoff energy than the measured value of Eq.\ \eq{eq:cutoff_K} and considerably increases the longitudinal and transverse responses of the computed $p(\gamma^*,K^+)\Lambda$ cross sections. In a similar vein, the VGL model adopts \cite{Guidal:1999qi}
\begin{align}
\Lambda_{\gamma K K}^\text{VGL} = \Lambda_{\gamma K^* K}^\text{VGL} \simeq 1.2 \GeV .
\end{align}
Guidal {\it et al.\ }argue that for the $K$ this could be attributed to the fact that the pole in the kaon propagator $(t - m_K^2)^{-1}$ is further from the physical region, compared to the pion case $(t - m_\pi^2)^{-1}$. Hence, the high $\Lambda_{\gamma K K}$ value would be representative for the whole kaon-Regge trajectory, rather than for the physical kaon.

Figures \ref{fig:electro_JLab12} and \ref{fig:electro_JLab12-Q2} show the VR and RPR-2011 predictions for the $p(\gamma^*,K^+)\Lambda$ \textsc{l}-\textsc{t}-separation experiment planned for the 12 GeV upgrade at JLab \cite{Horn:2008pr}. From both figures it is clear that the VR model predicts both substantially smaller longitudinal and larger transverse cross sections than the RPR-2011 model. For the $\s{l}$ this can be mainly attributed to the adopted values of $\Lambda_{\gamma K K}$ and to a smaller extent of $\Lambda_{\gamma K^* K}$. In particular at small $-t$, where $t$-channel $K$ exchange is dominant, the magnitude of $\s{l}$ is very sensitive to the value of $\Lambda_{\gamma K K}$. On the other hand, the larger transverse response in the VR model can be attributed to the R-P contributions in the gauge-fixing $s$ channel. This is a key element of the VR framework and is not present in the RPR-2011 and the VGL model.

In $\pi^+n$ electroproduction, hadronic models like the VGL model cannot account for the anomalously large $\s{t}$ above the resonance region \cite{Blok:2008jy}. A similar scenario is expected in high-energy $K^+\Lambda$ electroproduction. Indeed, when adopting the experimental value of Eq.\ \eq{eq:cutoff_K} for $\Lambda_{\gamma K K}$, the VGL model, for example, significantly underpredicts the unseparated Cornell data shown in Fig.\ \ref{fig:electro_Cornell}. Also the VR model without R-P effects substantially underpredicts these data (Figs.~\ref{fig:electro_Cornell} and \ref{fig:electro_CLAS}). Given that $\s{u} = \s{t} + \varepsilon\s{l}$ and that $\Lambda_{\gamma K K}$ predominantly influences $\s{l}$ at forward scattering, the much larger kaon cutoff energy required by the VGL model might actually be a compensation for an increased transverse response which remains unrevealed in $\s{u}$. In this respect, the VR framework constitutes a promising approach as it inherently accounts for a larger $\s{t}$ and already successfully explains the separated structure functions, measured in high-energy pion electroproduction \cite{Vrancx2013}.

The JLab \textsc{l}-\textsc{t}-separation experiment for high-energy $K^+\Lambda$ electroproduction is expected to settle the magnitude of the $\s{t}$ response. In addition, the measurement of $\s{l}$ at small $-t$ will provide access to the value of $\Lambda_{\gamma K K}$ in off-shell circumstances. Another experiment is planned with the 12 GeV upgrade at JLab. The CLAS Collaboration intends to obtain the interference structure functions $\s{lt}$, $\s{tt}$, and $\s{lt'}$ for $Q^2$ and $W$ values up to $12\GeV^2$ and $3\GeV$ \cite{Carman:2014}. These data will also constitute an important test bed for the VR model as the proposed kinematics cover the trans-resonance region.

%######################################################################
\section{Conclusion and outlook}
\label{sec:conclusions}
Building on the VR model for charged-pion electroproduction, the VR model for the $p(\gamma^*,K^+)\Lambda$ reaction above the resonance region and forward angles ($-t < 1\GeV^2$) was introduced. This model uses a three-trajectory Regge model for the photoproduction reaction as a starting base. The model features one pseudoscalar- and two vector-kaon Regge trajectories in the $t$ channel. It provides an excellent description of the high-energy ($W > 2.6 \GeV$), forward-angle ($\cos\thcm > 0.35$) cross section and recoil polarization $p(\gamma,K^+)\Lambda$ data from the CLAS Collaboration. Turning to finite photon virtualities, a key feature of the VR model for $p(\gamma^*,K^+)\Lambda$ is to introduce a proton EM transition form factor, accounting for the contributions of resonances-partons connected to the highly off-shell proton in the gauge-fixing $s$-channel. The same proton transition form factor is assumed in both $\pi^+n$ and $K^+\Lambda$ electroproduction. The magnitude of the antishrinkage effect in the $s$-channel is the sole parameter of the VR model and was optimized against the scarce $p(\gamma^*,K^+)\Lambda$ data. Remarkably, its optimized value coincides with the one obtained in pion electroproduction, for which far more data is available.

After introducing the R-P contributions, a good theory-experiment agreement is achieved for the 19 unseparated $p(\gamma^*,K^+)\Lambda$ cross-section data $\s{u}$ for $W > 2.5 \GeV$ and $-t \lesssim 1\GeV^2$. To date, only six data points are available for the interference structure functions. Due to limited statistics, the situation is rather inconclusive for those. An alternate explanation of the anomalous magnitude of the measured $\s{u}$ is that the kaon electromagnetic form factor in $t$-channel $p(\gamma^*,K^+)\Lambda$ is substantially harder than in elastic $eK$ scattering. 

Predictions are provided for the upcoming \textsc{l}-\textsc{t}-separation experiment at JLab. This experiment will provide the first data for the $p(\gamma^*,K^+)\Lambda$ longitudinal and transverse responses above the resonance region. It is expected that the forward-scattering $\s{l}$ data will map the kaon electromagnetic form factor in $K^+\Lambda$ electroproduction. The $\s{t}$ data, on the other hand, will reveal the importance of additional model features, like the role of resonance-parton effects. In high-energy pion electroproduction these provide a natural explanation for the observed magnitude of the transverse response. It is to be awaited whether or not this is the case in kaon electroproduction.

%######################################################################
\acknowledgments
This work is supported by the Research Council of Ghent University and the Flemish Research Foundation (FWO Vlaanderen). The authors would like to thank Daniel Carman and Viktor Mokeev for providing the recent $K^+\Lambda$ electroproduction data from the CLAS Collaboration and for useful discussions.

%######################################################################
\appendix*
\section{Observables}
\label{sec:app_observables}

\subsection{Photoproduction}
\label{subsec:app_photo}
The laboratory frame coordinate system is defined as
\begin{align}
\bm{z} = \frac{\bm{q}}{|\bm{q}|}, \quad \bm{y} = \frac{\bm{q}\times\bm{k}'}{|\bm{q}\times\bm{k}'|}, \quad \bm{x} = \bm{y}\times\bm{z},
\end{align}
where $\bm{q}$ and $\bm{k}'$ are the three-momenta of the photon and outgoing kaon. The hadronic matrix elements $M^\lambda_{\lambda_p,\lambda_\Lambda}$ are defined as
\begin{align}
M^\lambda_{\lambda_p, \lambda_\Lambda} = \epsilon^\lambda_\mu J^\mu_{\lambda_p, \lambda_\Lambda},
\end{align}
with $J^\mu_{\lambda_p, \lambda_\Lambda}$ being the transition current of the $p(\gamma^{(*)},K^+)\Lambda$ reaction and $\epsilon^\lambda_\mu$ the covariant polarization four-vector of the $\gamma^{(*)}$. For the observables covered in this work, it suffices to consider circularly polarized photons:
\begin{align}
\epsilon^\pm_\mu = \frac{1}{\sqrt{2}}(0,\pm 1,i,0).
\end{align}
The unpolarized $p(\gamma,K^+)\Lambda$ differential cross section is calculated as
\begin{align}
\frac{\d\sigma}{\d\Omega_K^*} = \frac{\alpha_e m_pm_\Lambda|{\bm{k}'}^*|}{16\pi W(s - m_p^2)}\sum_{\lambda,\lambda_p,\lambda_\Lambda}\bigl|M^\lambda_{\lambda_p, \lambda_\Lambda}\bigr|^2,
\end{align}
where
\begin{align}
|{\bm{k}'}^*| = \sqrt{\frac{(s + m_K^2 - m_\Lambda^2)^2}{4s} - m_K^2},
\end{align}
is the size of the three-momentum of the outgoing kaon in the center-of-mass frame. In terms of the hadronic matrix elements, the recoil polarization $P$ reads
\begin{align}
P = \frac{\sum_{\lambda,\lambda_p}\Bigl(\bigl|M^\lambda_{\lambda_p, \lambda_\Lambda=+y}\bigr|^2 - \bigl|M^\lambda_{\lambda_p, \lambda_\Lambda=-y}\bigr|^2\Bigr)}{\sum_{\lambda,\lambda_p,\lambda_\Lambda}\bigl|M^\lambda_{\lambda_p, \lambda_\Lambda}\bigr|^2}.
\end{align}

\subsection{Electroproduction}
\label{subsec:app_electro}
In electroproduction, the photon is virtual and a longitudinal polarization is allowed:
\begin{align}
\epsilon^0_\mu = \frac{1}{\sqrt{Q^2}}(\sqrt{\nu^2 + Q^2},0,0,-\nu),
\end{align}
with $\nu = E_e - E_{e'}$ being the energy difference between the initial and final electrons, $e$ and $e'$. The unseparated differential cross section reads
\begin{align}
\frac{\d\s{u}}{\d t} = \frac{\d\s{t}}{\d t} + \varepsilon\frac{\d\s{l}}{\d t},\label{eq:sigma_unseparated}
\end{align}
with $\varepsilon$ given by
\begin{align}
\varepsilon = \frac{4E_eE_{e'} - Q^2}{2(E_e^2+E_{e'}^2) + Q^2}.
\end{align}
The longitudinal and transverse structure functions are calculated as
\begin{align}
\frac{\d\s{l}}{\d t} &= 2\eta H_{0,0},\nonumber\\
\frac{\d\s{t}}{\d t} &= \eta(H_{+,+} + H_{-,-}), \label{eq:sigma_separated}
\end{align}
where
\begin{align}
H_{\lambda, \lambda'} = \sum_{\lambda_p,\lambda_\Lambda} M^\lambda_{\lambda_p,\lambda_\Lambda} \Bigl(M^{\lambda'}_{\lambda_p,\lambda_\Lambda}\Bigr)^*.
\end{align}
The normalization factor $\eta$ reads
\begin{align}
\eta = \frac{\alpha_e m_p m_\Lambda}{4W(s - m_p^2)|\bm{q}^*|},
\end{align}
with
\begin{align}
|\bm{q}^*| = \sqrt{\frac{(m_p^2 - s + Q^2)^2}{4s} + Q^2},
\end{align}
being the size of the photon's three-momentum in the center-of-mass frame. The interference structure functions for the longitudinal and transverse components of the virtual photon polarization can be expressed as
\begin{align}
\frac{\d\s{lt}}{\d t} &= -\eta(H_{+,0} + H_{0,+} - H_{0,-} - H_{-,0}),\nonumber\\
\frac{\d\s{tt}}{\d t} &= -\eta(H_{+,-} + H_{-,+}),\nonumber\\
\frac{\d\s{lt'}}{\d t} &= -\eta(H_{+,0} - H_{0,+} + H_{0,-} - H_{-,0}).
\label{eq:sigma_interference}
\end{align}
Finally, the transformation from $\d t$ to $\d\Omega_K^*$ can be accomplished by employing the relation
\begin{align}
\frac{\d\Omega_K^*}{\d t} = \frac{\pi}{|\bm{q}^*||{\bm{k}'}^*|}.\label{eq:dOmegadt}
\end{align}
Note that, strictly speaking, the differential ``$\d t$'' in Eqs.\ \eq{eq:sigma_unseparated}, \eq{eq:sigma_separated}, \eq{eq:sigma_interference} and \eq{eq:dOmegadt} should read ``$-\d t$''. It is a conventional, however, to write ``$\d t$'' in the expressions for the differential cross sections.

%######################################################################

\end{document}